\documentclass[aps,prl,twocolumn,epsfig,showpacs]{revtex4}
\usepackage{dcolumn}
\usepackage{bm}
\usepackage{graphicx}
\usepackage{amsmath}
\usepackage{latexsym}
\usepackage{amsfonts}
\usepackage{amssymb}
\usepackage{array}
\usepackage{epsfig}
\usepackage{epstopdf}
\newcommand{\ket}[1]{\left\vert#1\right\rangle}
\newcommand{\bra}[1]{\left\langle#1\right\vert}

\begin{document}
\title{
Failure of Local Realism Revealed by Extremely Coarse-Grained Measurements
}
\author{
Hyunseok Jeong,$^{1,2}$ Mauro Paternostro,$^3$ and Timothy C. Ralph$^1$}
\affiliation{
$^1$Centre for Quantum Computer Technology,
Department of Physics,
University of Queensland, St Lucia, Qld 4072, Australia\\
$^2$Center for Subwavelength Optics and 
Department of Physics and Astronomy,
Seoul National University,
 Seoul, 151-742, Korea\\
$^3$School of Mathematics and Physics, The Queen's University, Belfast,
BT7 1NN, UK}
\date{\today}

\begin{abstract}
We show that failure of local realism
can be revealed to observers for whom only extremely coarse-grained measurements are available.
In our instances, Bell's inequality is violated even up to the maximum limit while
both the local measurements and the initial local states under scrutiny approach the classical limit.
Furthermore, we can observe failure of local realism when
 an inequality enforced by non-local realistic theories is satisfied.
This suggests that
locality alone may be violated while realism cannot be excluded for
specific observables and states.  Small-scale experimental
 demonstration of our examples may be possible in the foreseeable future.
\end{abstract}
\pacs{03.67.Mn, 42.50.Dv, 03.65.Ud, 42.50.-p}

\maketitle

The development of quantum physics has revealed
a world quite different from the one depicted by classical physics.
Probably, the most striking feature of the quantum world,
distinguishing it from the classical one,
is failure of local realism~\cite{EPR,Bell}.
Local realism combines two reasonably acceptable
assumptions, locality and realism.
The principle of locality is that
distant objects cannot have direct instantaneous influence on one another.
Physical realism
claims that all measurement outcomes are
determined by pre-existing quantities of physical systems.
The failure of local realism is evidenced by
violation of Bell's famous inequality which should be obeyed by
any local-realistic theories.

Although certain odd features of nature predicted by quantum physics
such as the failure of local realism
have been experimentally observed in laboratories~\cite{exp1,exp2},
such quantum properties are not seen in our everyday experience on a macroscopic scale.
Decoherence is often considered the main reason for
the appearance of the classical world from the laws of quantum physics \cite{decoherence}.
Quantum systems, particularly when they are macroscopic,
unavoidably interact with their environments and rapidly lose their quantum features.
Recently, Kofler and Brukner
suggested a conceptually different view~\cite{Kofler}:
 they attributed the appearance of the classical
world to the ``coarse-grained" (or fuzzy) properties of the measurements,
suggesting that ``classical physics can
be seen as implied by quantum mechanics under the restriction
of fuzzy measurements.''

Here, we address a crucial question concerning fundamental tests of quantum mechanics:
Can the quantum world where local realism fails be perceived by the observer
even when the measurements are very unsharp?
In stark contrast with the conclusions reached in~\cite{Kofler}, here we find that
extremely coarse-grained measurements can still be useful to reveal
the quantum world where local realism fails.
In our examples, Bell's inequality, which is enforced by local realism,
is violated up to the maximum limit, known
as Cirel'son's bound~\cite{C80}, when appropriate local unitary transformations and states are chosen
alongside the coarse-grained measurements.
Furthermore, we show that while local realism fails,
a recent version of Leggett's inequality~\cite{Zeilinger07b,Leggett,Zeilinger07},
which is imposed by non-local realistic theories, can still be satisfied.
Failure of local realism means that at least one between locality and realism
should be abandoned while the satisfaction of Leggett's inequality implies that
a realistic interpretation is tenable as far as some level of non-locality is allowed.
This suggests that locality alone fails while realism is tenable in our specific examples.

In order to show such instances,
we first need to describe the sort of coarse-grained measurements we consider, which
 should only be able to discern differences at a macroscopic scale~\cite{Kofler}.
As an extreme example, a ``measurement" by human eyes can notice differences between two objects
only when they are macroscopically different.
In quantum optics, homodyne measurements with low efficiency
can be considered such coarse-grained measurements in the classical limit:
two different states may be distinguished only when they
are sufficiently separate in phase space.
In our study, we make use of
entangled thermal states (ETSs), which have been introduced in~\cite{jr06},
where component states are ``classical'' thermal states.
Of course, even when local states obtained by taking the partial trace of the total
state appear classical, Bell's inequality can be violated if ``sharp''
measurements, such as highly efficient photon number detection,
are used as shown in \cite{jr06,added}. However, it has not been previously found that
extremely unsharp measurements can be used to reveal failure of local realism.

In our proposal, two local parties, Alice and Bob, are each provided
with one mode of an ETS prepared by a third party upon entangling
two displaced thermal states as described in \cite{jr06}.
A displaced thermal state is defined as
$\rho^{\rm th}(V,d)=\int d^2\alpha P_\alpha^{\rm th}(V,d)
|\alpha\rangle\langle\alpha|$,
where
$P_\alpha^{\rm th}(V,d)=\frac{2}{\pi(V-1)}
e^{-\frac{2|\alpha-d|^2}{V-1}}$ a Gaussian function
with variance $V$ and center
$d$ (with respect to the origin of the phase space).
Two identical such states, $\rho^{\rm th}_A(V,d)$ and
$\rho^{\rm th}_B(V,d)$, are distributed to spatially separate
locations. As the first step to entangle them,
a microscopic system in the superposition state,
$|\psi\rangle_m=(|0\rangle_m+|1\rangle_m)/\sqrt{2}$,
sequentially
interacts with the two thermal states,
where, $|0\rangle_m$ and $|1\rangle_m$ are the ground and excited states
of the microscopic system (for instance, the first two levels of a
harmonic oscillator or two energy levels of an atom).
The interaction is taken to be of the nonlinear cross-Kerr form
${\cal H}^K_{mj}= \hbar\lambda \hat a_m^\dagger \hat a_m \hat a_j^\dagger
\hat a_j$ with $\lambda$ the strength of the nonlinearity and $j=A,B$.
Nonlinear media with free-traveling optical fields~\cite{free1,free2}
or dispersive interactions within optical/microwave cavities~\cite{cavities}
may be used to implement such interactions~\cite{jr06,paternostro06}.
We stress that the use of the microscopic superposition,
$|\psi\rangle_m$, is {\it not} essential, and we shall later describe
an alternative method using another type of ETS produced without it.

For simplicity, we assume that the interaction time is
$t=\pi/\lambda$, while we note that an equivalent effect can be obtained
in principle using a (more realistic) weaker nonlinearity ($t\ll\pi/\lambda$)
and a thermal state with a larger displacement $d$~\cite{free2,jr06}.
When the thermal state $\rho^{\rm th}_j(V,d)$ interacts with the ground
state $|0\rangle_m$, nothing happens. On the other hand,
when it interacts with the excited state of $m$,
it evolves to $\rho^{\rm th}_j(V,-d)$, {\it i.e.}
it is ``moved" to the opposite location in the phase space.
After the interactions represented by
${\cal H}^K_{mA}\otimes{\cal H}^K_{mB}$,
the microscopic system is measured out on the superposed basis
$(|0\rangle\pm|1\rangle)_m/\sqrt{2}$.
As the result, the thermal states at modes $A$ and $B$ become
entangled~\cite{jr06,paternostro06}.
From now on, we assume that the outcome of the measurement is
$(|0\rangle+|1\rangle)_m/\sqrt{2}$ so that the ETS $\rho_{AB}^{\Psi(+)}$
is shared between Alice and Bob who, as described below, should now perform 
nonlinear local operations and homodyne measurements.

The local operations required for Bell inequality tests
are composed of displacement operation $\hat
D_j(\zeta)=e^{\zeta\hat{a}^{\dagger}_j-\zeta^{*}\hat{a}_j}$
($\zeta\in\mathbb{C}$)
and single-mode Kerr nonlinearities described by the interaction Hamiltonian
${\cal H}_{NL,j}= \hbar\Omega({\hat a}^\dag_j {\hat a}_j)^2$,
where $\Omega$ is the strength of the nonlinearity
and $\hat{a}_j$ ($\hat{a}^\dagger_j$) is the
annihilation (creation) operator of system $j$.
The displacement operation $\hat D_j(\zeta)$
can be readily performed using a beam splitter and
a thermal state with a large displacement.
Nonlinear media such as optical crystals can in principle be used to
realize single-mode Kerr effects.
It is known that the Kerr nonlinear interaction,
$\hat U_{NL}=e^{-\frac{i}{\hbar} {\cal H}_{NL}t_c}$,
where $t_c=\pi/\Omega$, causes a coherent state to evolve into the normalized
state $\hat U_{NL}|\alpha\rangle
=e^{-i\frac{\pi}{4}}(|\alpha\rangle+i|-\alpha\rangle)/\sqrt{2}$ \cite{YS86}.
We define the local unitary operations~\cite{SJR}
$\hat {\cal V}_j(\theta_j) = \hat U_{NL,j} \hat D_j\left(
{i\theta_j}/{2d}\right) \hat U_{NL,j}$, which are applied to mode $j=A,B$ as
${\rho_{AB}^{\Psi{(+)}}}^{\prime}(\theta_A,\theta_B)=
\hat {\cal V}_A(\theta_A)\hat {\cal V}_B(\theta_B)
{\rho_{AB}^{\Psi{(+)}}}
\hat {\cal V}^\dagger_A(\theta_A)\hat {\cal V}^\dagger_B(\theta_B)$.

An imperfect homodyne detector with efficiency $\eta$
can be modeled by a beam splitter with transmittivity $\eta$,
superimposing mode  $j$ ($j=A,B$) with an ancilla $v_j$ prepared in vacuum state,
and cascaded with an ideal homodyne detector.
 The beam splitter operator between
modes $j$ and $v_j$ is defined as
${\hat B}_{jv_j}=e^{\frac{\xi}{2}({\hat a}_j^\dagger {\hat a}_{v_j}
-{\hat a}_j {\hat a}_{v_j}^\dagger)}$,
where $\cos\xi=\sqrt{\eta}$. As we discard the output state of the
ancillae, this changes
${\rho_{AB}^{\Psi{(+)}}}^{\prime}(\theta_A,\theta_B)$ into
${\rho_{AB}^{\Psi{(+)}}}^{\prime}(\theta_A,\theta_B,\eta)
={\rm Tr}_{v_{A,B}}\big[{\hat B}_{Av_A}{\hat
B}_{Bv_B}{\rho_{AB}^{\Psi{(+)}}}^\prime(\theta_A,\theta_B)
\big(|00\rangle_{v_A v_B}\!\langle{00}|\big){\hat B}^\dagger_{Av_A}{\hat
B}^\dagger_{Bv_B}\big]$.

\begin{figure}[b]
{(a)}\hskip2cm{(b)}\hskip2cm{(c)}
\centerline{\scalebox{0.3}{\includegraphics{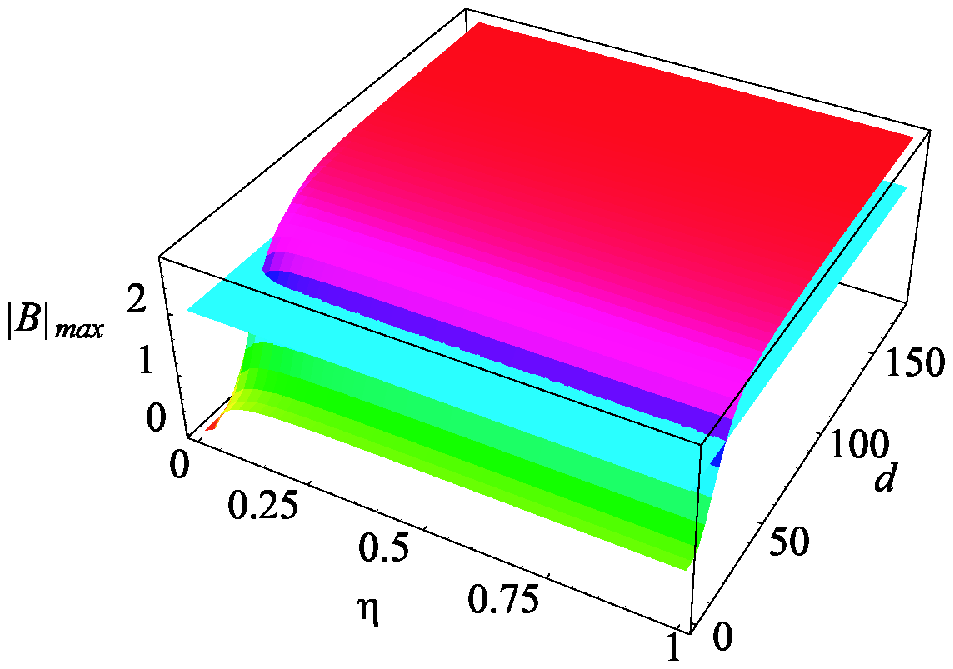}}~\hskip-0.05cm
\scalebox{0.3}{\includegraphics{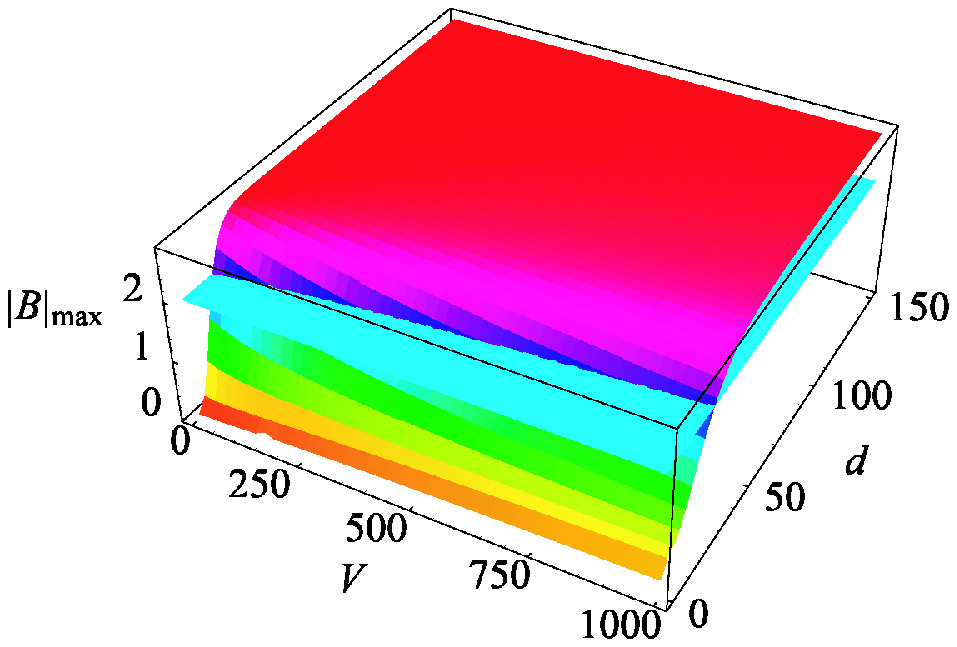}}~\hskip-0.05cm
\scalebox{0.207}{\includegraphics{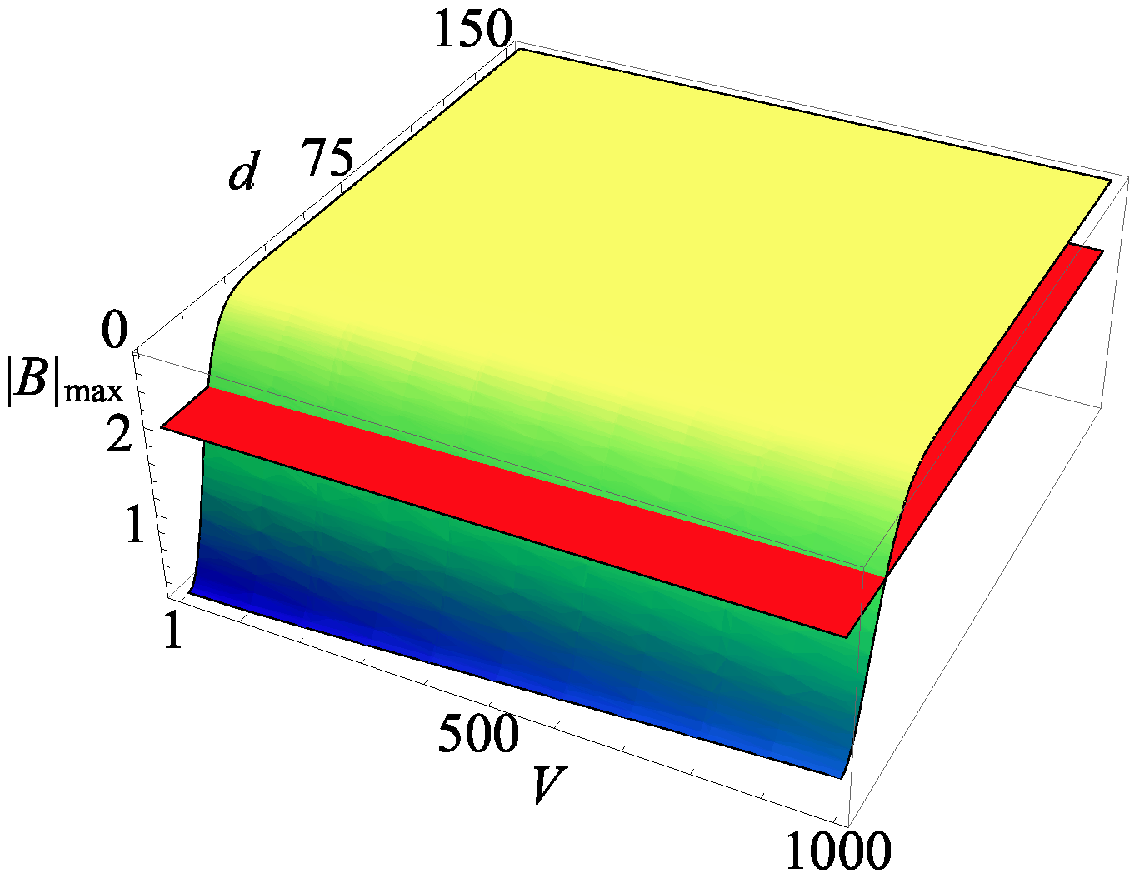}}}
\caption{
(a) The numerically optimized Bell function $|B|_{\rm max}$
for the considered ETS
as a function of the displacement $d$
and homodyne efficiency $\eta$
for $V=1000$.
(b) The Bell function $|B|_{\rm max}$
as a function of $d$
and the variance of the initial thermal state $V$
when $\eta=0.05$.
(c) Same as panel (b) but for the
alternative form of ETS $\rho^{\rm{alt}}_{AB}$ defined in the manuscript.
The horizontal plane in each figure indicates
the classical limit 2.}
\label{fig:3d}
\end{figure}

The tools described above bear some analogies
with the qubit case: states $\rho^{\rm th}(V,\pm d)$
correspond to a qubit basis, the ETS a two-qubit entangled state,
the nonlinear operation ${\cal V}_j(\theta_j)$ a single-qubit operation,
and the homodyne detection  to discriminate between the
two thermal states a computational basis measurement.
In order to test Clauser, Horne, Shimony, and Holt (CHSH)'s
version of Bell's inequality~\cite{CHSH},
we assign value $+1$ to the measurement outcome
corresponding to a homodyne signal larger than 0,
and $-1$ otherwise.
Then, the
{\it joint} probability
$P_{kl}(\theta_A,\theta_B)$,
where the subscripts $k,l=\pm$ correspond to A and B's
assigned measurements outcomes $\pm{1}$ respectively,
can be calculated as
$P_{kl}(\theta_A,\theta_B)=
\int^{k_s}_{k_i} dx \int^{l_s}_{l_i} dy~_A\langle x|_B\langle y|
{\rho_{AB}^{\Psi{(+)}}}^{\prime}(\theta_A,\theta_B,\eta)|x\rangle_A|y\rangle_B$,
where $|x\rangle$ and $|y\rangle$ are quadrature eigenstates.
The Bell function is constructed as
$B(\theta_A,\theta_B,\theta^\prime_A,\theta^\prime_B)=
C(\theta_A,\theta_B)+C(\theta_A^\prime,\theta_B)+
C(\theta_A,\theta^\prime_B)-C(\theta^\prime_A,\theta^\prime_B)$
where $C(\theta_A,\theta_B)=\sum_{k=\pm}P_{kk}(\theta_A,\theta_B)
-\sum_{k\neq{l}=\pm}P_{kl}(\theta_A,\theta_B)$
and the Bell-CHSH inequality~\cite{CHSH} is
$|B(\theta_A,\theta_B,\theta^\prime_A,\theta^\prime_B)|\leq2$. Throughout this
process, we have obtained the Bell's function
$B(\theta_A,\theta_B,\theta^\prime_A,\theta^\prime_B)$
as a function of $V$, $d$ and $\eta$.
The explicit form of $C(\theta_A,\theta_B)$ which composes
the Bell's function is
\begin{equation}
\begin{split}
&C(\theta_A,\theta_B,\eta)=
YW\{e^{4i\theta_A}g(\theta_A)
[ie^{\frac{2d^2}{V}}Yh(\theta_B)\kappa(\theta_B,\theta_B)\\
&+Qg(\theta_B)s_{\theta_B}]\!+\!Yh(\theta_A)[ie^{4i\theta_B
+2\frac{V\theta^2_B}{d^2}}g(\theta_B)
s(\theta_B)\kappa(\theta_A,\theta_B)\\
&+4Yh(\theta_B)\left(e^{8i\theta_A}f_-(\theta_B)f_+(\theta_A)
+e^{8i\theta_B}f_-(\theta_A)f_+(\theta_B)\right)]\}
\end{split}
\end{equation}
Here we have defined $Y=[8(1+V^2e^\frac{4d^2}{V})]^{-1}$,
$h(\mu)=e^\frac{2(d^4+\mu^2)}{d^2V}$,
$W=e^{-4i(\theta_A+\theta_B)-\frac{2(1+V^2)(\theta^2_A+\theta^2_B)}{d^2V}}$,
$Q=8e^{4i\theta_B+\frac{2V(\theta^2_A+\theta^2_B)}{d^2}}$,
$f_\pm(\mu)={\rm Erf}\Big(\frac{\sqrt{2}\eta(d^2\pm i V\mu)}
{d\sqrt{1+\eta^2(V-1)}}\Big)$, $\kappa(\mu,\nu)=f_-(\mu)-e^{8i\nu}f_+(\mu)$,
$g(\mu)={\rm Erfi}\Big(\frac{\sqrt{2}\eta\mu}
{d\sqrt{V^2-\eta^2V(V-1)}}\Big)$ and $s_\mu={\rm sign(\mu)}$ with
$\mu,\nu=\theta_A,\theta_B$.
We have then numerically optimized the Bell's function to obtain
$|B(\theta_A,\theta_B,\theta^\prime_A,\theta^\prime_B)|_{\rm max}$, which is plotted
in Figs.~\ref{fig:3d} against the relevant parameters $V$, $d$ and $\eta$.

In Fig.~\ref{fig:3d}(a), large thermal states with
$V=1000$ have been used in order to generate the various ETSs.
The (sky-blue) horizontal plane
indicates the classical limit 2 over which
local-realistic theories fail.
It is evident that the Bell-CHSH inequality is significantly
violated for large regions, almost uniformly with respect to $\eta$.
In this example, the degree of mixedness for the ETS $\rho^{\Psi(+)}$,
quantified by the linear entropy $S(V,d)=1-{\rm Tr}[{{\rho^{\Psi(+)}_{AB}}}^2]$, is
$S(1000,d)>0.999$, regardless of $d$.
Here, $S(V,d)=0$ for pure states while $S(V,d)=1$ for completely mixed ones.
In spite of nearly the maximum degree of mixedenss,
significant violations of Bell-CHSH inequality
are revealed.
In Fig.~1(b), we choose $\eta=0.05$, which is a very low detection
efficiency,
and again, strong violations are observed as $d$ increases.
In other words, the Bell's function rapidly approach
the maximum bound $2\sqrt{2}$, known as Cirel'son's bound~\cite{C80}
as $d$ increases, regardless of values of $\eta$ and $V$.
We have therefore shown that the Bell-CHSH inequality can
be violated nearly up to the maximum value even when extremely coarse-grained
measurements are used, as far as one can increase the ``classical distinctness"
$d$ between the local states.

It is worth noting that in a sense, a ``more classical'' method without the microscopic superposition
results in qualitatively the same conclusions: 
an alternative form of ETS is given by superimposing a displaced therma
mode $A$ subjected to a  single-mode Kerr nonlinear interaction
$\hat{\cal H}_{NL,A}$ with a vacuum mode $B$ at a $50\!:\!50$ beam splitter.
The resulting state has the structure 
$\rho^{\rm{alt}}_{AB}=\int{d}^2\alpha{P}^{\rm{th}}_{\alpha}(V,d)\ket{\psi}_{AB}\!\bra{\psi}$
 with $\ket{\psi}_{AB}\propto\ket{{\alpha}\sqrt{2},-\alpha/\sqrt{2}}_{AB}
 +i\ket{-\alpha/\sqrt{2},\alpha/\sqrt{2}}_{AB}$.
Here, in order to show Bell violations in the same way as before,
the third party needs to perform an additional displacement
operation $\hat{D}_A(i\pi/8d)$ to remove the relative phase factor, $i$,
of the alternative form of ETS that was not present in the previously discussed case.
The state can then be shared by Alice and Bob for the Bell inequality test,
and the construction of the corresponding Bell function follows the steps described above.
Although this time we have not been able to get a closed analytical expression for
$C(\theta_A,\theta_B,\eta)$,
a numerical calculation reveals qualitatively the same features
to those found for the case of $\rho^{\Psi(+)'}_{AB}$, as shown in Fig.~1(c).
Without loss of generality, we restrict our study to the
$\rho^{\Psi(+)'}_{AB}$ class of states 
due to the convenience of dealing with fully analytic expressions. Of course, our results
can be extended to $\rho^{\rm{alt}}_{AB}$.

A natural question now arises: ``What
causes Bell's inequality to be violated even when
local states and measurements are totally classical''?
We stress that single-mode Kerr nonlinear interactions are important elements of
the local operation, $\hat {\cal V}_j(\theta_j)$.
It is straightforward to show that the dynamics of the thermal state
in a Kerr medium, which can be observed by homodyne measurements,
differs from that of the classical counterpart.
The latter can be obtained by replacing quantum mechanical
operator $\hat a$ ($\hat a^\dagger$) with $c$-number $\alpha$
($\alpha^*$)~\cite{Walls}.
In other words, Alice and Bob can independently observe
statistical distributions of homodyne measurement results,
even when the detection efficiency is low, different from the
distributions predicted by the classical theory of light in a Kerr medium.
Therefore, it is still questionable whether one can find a realistic description
to explain all the measurement results for each local system.
To pursue an answer to this question,
we investigate non-local realism as expressed by a recent version of
Leggett's inequality~\cite{Leggett,Zeilinger07b}. Following the derivation
given in~\cite{Zeilinger07b}, one finds that $\hat{R}_j(\theta_j,\varphi_j)
=\left(\begin{array}{cc}\sin(\theta_j/2)&e^{i\varphi_j}\cos(\theta_j/2)\\
e^{-i\varphi_j}\cos(\theta_j/2)&-\sin(\theta_j/2)\end{array}\right)$, applied
to the vector $\begin{pmatrix}\ket{\alpha}_j&\ket{-\alpha}_j\end{pmatrix}^T$,
realizes the set of local operations needed for this task. Notice that such
a set requires out-of-plane rotations. We have thus generalized our effective
transformations by following the scheme in~\cite{jacobandmyung}, through which
one can recognize that the sequence
$\hat{D}_j(-i\varphi_j/4\alpha)\hat{U}_{NL}\hat{D}_j
(i\theta_j/4\alpha)\hat{U}_{NL}\hat{D}_j(i\varphi_j/4\alpha)$ 
approximates $\hat{R}_j(\theta_j,\varphi_j)$. From now on, 
we identify such operations by specifying the unit
vectors $\mathbf{a}\equiv(\theta_A,\varphi_A)$ and
$\mathbf{b}\equiv(\theta_B,\varphi_B)$, determined by the corresponding
set of angles expressed in spherical polar coordinates.
Then, we identify the unit vectors $\mathbf{a}\equiv(\theta_A,\varphi_A)$
and $\mathbf{b}\equiv(\theta_B,\varphi_B)$ by the set of corresponding
angles in spherical polar coordinates. By
following the procedure described for the Bell-CHSH inequality
and through a rather lenghty calculation, one
can find the form of the correlation function $C^L_{}(\mathbf{a},\mathbf{b})$
associated with non-ideal detectors. However, the expression
we gather is too lengthy to be reported here and we thus directly pass
to discuss our results.

\begin{figure}
\centerline{\scalebox{0.5}{\includegraphics{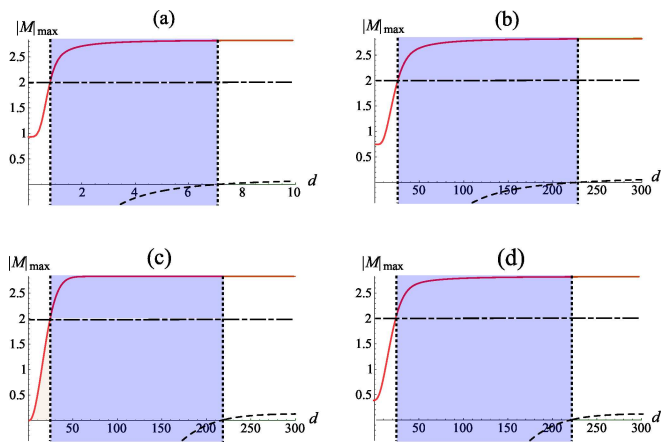}}}
\caption{
The optimized Bell function $|B|_{\rm max}$ and the
optimized Leggett function $|{\cal L}|_{\rm max}$
are presented, where $M=B$ for the solid curve
and $M={\cal L}$ for the dashed curve.
(a) When $V=1$ and $\eta=1$ (pure entangled coherent states and
perfect detectors), there is a range of $d$
where the Bell-CHSH inequality is violated while the Leggett's inequality
is satisfied.
(b) When $V=1000$ (highly mixed ETSs), this range becomes larger.
(c) The same effect can be obtained by decreasing $\eta$ to $0.03$.
(d) An example with $V=700$ and $\eta=0.05$ is given.}
\label{fig:leggett}
\end{figure}

Non-local realism can be studied by considering the unit vectors
$\mathbf{a}_{1,2,3}$ and $\mathbf{b}_{1-7}$, each
identifying a rotation that A (B) has to perform on her (his) mode.
Explicitly $\mathbf{a}_{1}\!=\!\mathbf{b}_{5}\equiv({\pi}/{2},0)$,
$\mathbf{a}_{2}\!=\!\mathbf{b}_{6}\equiv({\pi}/{2},{\pi}/{2})$,
$\mathbf{a}_{3}\!=\!\mathbf{b}_{7}\equiv({0}{},{0}{})$,
$\mathbf{b}_{1}\equiv({\pi}/{2},\varphi)$ and
$\mathbf{b}_{4}\equiv({\varphi},{\pi}/{2})$ with $\mathbf{b}_{2}$ and
$\mathbf{b}_{3}$ which are found from $\mathbf{b}_{1}$ and
$\mathbf{b}_{4}$, respectively, by taking
$\varphi\rightarrow\pi/2+\varphi$.
We can thus build the function
\begin{equation}
\label{leggettine}
\begin{split}
L\!=&\!|C^L_{}(\mathbf{a}_1,\mathbf{b}_1)\!+\!C^L_{}
(\mathbf{a}_2,\mathbf{b}_2)\!+\!C^L_{}
(\mathbf{a}_1,\mathbf{b}_5)\!+\!C^L_{}(\mathbf{a}_2,\mathbf{b}_6)|\\
\!&+\!|C^L_{}(\mathbf{a}_2,\mathbf{b}_3)\!+\!C^L_{}
(\mathbf{a}_3,\mathbf{b}_4)\!+\!C^L_{}(\mathbf{a}_2,
\mathbf{b}_6)\!+\!C^L_{}(\mathbf{a}_3,\mathbf{b}_7)|.
\nonumber
\end{split}
\end{equation}
Non-local realistic theories impose a bound on $L$ given by is
$8-2|\sin(\varphi/2)|$~\cite{Zeilinger07b}. Numerically, we have found
that the Leggett function defined (for convenience) as ${\cal
L}=L-8+|\sin(\varphi/2)|$ is maximized for $\varphi\sim{0.2507}$ radians,
which is the value we assume in our calculations.
With our notation, ${\cal L}\leq 0$ is forced
by non-local realistic theories.
If the inequality is satisfied,
we may ``retain" non-local realistic theories
to explain all the measurement results
under our assumptions~\cite{referee}.
To see if this is the case, we have studied the optimized Leggett function
$|{\cal L}|_{\rm max}$ against $V$ and $d$.
Our analysis shows that, indeed,
there is a range of values of $d$ where the Bell-CHSH inequality is
violated while Leggett's one
is satisfied for any given value of $V$ and $\eta$, thus confirming our
expectations.
Interestingly, as seen in Fig.~\ref{fig:leggett}, this range widens
by going towards regimes of increasing classicality,
{\it i.e.} when $V$ ($\eta$) is increased (reduced).

We have shown that the quantum world, where local realism fails,
can be revealed by extremely coarse-grained measurements,
which is in stark contrast to previous findings~\cite{Kofler}.
Furthermore, Bell-CHSH inequality can be violated
while a Leggett-type inequality 
imposed by certain 
non-local realistic theories 
is satisfied. When $V$ and $d$ take moderate values,
the decoherence effects of the type of states under our consideration
remain in a reasonable range~\cite{deco}.
A small-scale experimental demonstration of our examples,
such as one in Fig.~\ref{fig:leggett}(a), may be realized.
In this case, the required ETSs become pure
entangled coherent states (i.e., $V\sim1$)
with $d\approx 1.1$.
Such states can be generated, using a beam splitter, from
superpositions of two coherent states
with $d\approx 1.6$, which were experimentally demonstrated in a recent
experiment~\cite{sqcat}.
There have been important progresses~\cite{strong} in obtaining
strong nonlinearities, which are demanding yet
necessary to implement the local operation $\hat{\cal V}_j(\theta_j)$.
Our results unveil unknown aspects of
the boundary between quantum and classical worlds and
their small-scale experimental realization, although demanding, is 
foreseeable.

We thank A.C.~Doherty, Ph.~Grangier and M.S.~Kim for discussions.
This work was supported by the DTO-funded U.S. Army Research Office
(W911NF-05-0397), the ARC, the Queensland State Government,
and the KOSEF grant funded by the Korea government (MEST) (R11-2008-095-01000-0).
MP is supported by EPSRC (EP/G004579/1).


\begin{thebibliography} {99}


\bibitem{EPR} A. Einstein, B. Podolsky, and N. Rosen,
Phys.Rev. {\bf 47},
777 (1935).

\bibitem{Bell} J.S. Bell, Physics {\bf 1},
195 (1964).

\bibitem{exp1} S.J. Freedman and J.F. Clauser,
Phys.Rev.Lett. {\bf 28}, 938 (1972).

\bibitem{exp2} A. Aspect, Ph. Grangier and G. Roger,
Phys.Rev.Lett. {\bf 47}, 460 (1981).

\bibitem{decoherence} W.H. Zurek, Phys. Today {\bf 44}, 36 (1991); Rev.Mod.Phys. {\bf 75}, 715 (2003).

\bibitem{Kofler} J. Kofler and {\v C}. Brukner,
Phys.Rev.Lett. {\bf 99}, 180403 (2007).

\bibitem{C80} B.S. Cirel'son,
Lett.Math.Phys. {\bf 4}, 93 (1980).

\bibitem{Zeilinger07b}
T. Paterek {\it et al.},
Phys.Rev.Lett. {\bf 99}, 210406 (2007); C. Branciard {\it et al.}, {\it ibid.} {\bf 99}, 210407 (2007);
 C. Branciard {\it et al.}, Nature Phys. {\bf 4}, 681 (2008).

\bibitem{Leggett}
A.J. Leggett, Found. Phys. {\bf 33}, 1469 (2003).

\bibitem{Zeilinger07}
S. Gr\"oblacher {\it et al.},
Nature (London) {\bf 446}, 871 (2007).

\bibitem{jr06} H. Jeong and T.C. Ralph,
Phys.Rev.Lett. {\bf 97}, 100401 (2006);
Phys.Rev.A {\bf 76}, 042103 (2007).

\bibitem{added}
R. Filip, {\it et al.}, Phys.Rev.A {\bf 65}, 043802 (2002);
K. Banaszek and K. W\'odkiewicz, {\it ibid.} {\bf 58}, 4345
(1998);  Z.-B. Chen, {\it et al.},
Phys.Rev.Lett. {\bf 88}, 040406 (2002).

\bibitem{paternostro06} M. Paternostro, H. Jeong, and M.S. Kim,
Phys.Rev.A {\bf 73}, 012338 (2006).

\bibitem{free1} C.C. Gerry,
Phys.Rev.A {\bf 59}, 4095 (1999).

\bibitem{free2} H. Jeong,
Phys.Rev.A {\bf 72}, 034305 (2005).

\bibitem{cavities} M. Brune, {\it et al.}
Phys.Rev.Lett. {\bf 77}, 4887--4890 (1996).

\bibitem{YS86} B. Yurke and D. Stoler, Phys.Rev.Lett. {\bf 57}, 13 (1986).

\bibitem{SJR} M. Stobi\'nska, H. Jeong, and T.C. Ralph,
Phys.Rev A {\bf 75}, 052105 (2007).

\bibitem{CHSH} J.F. Clauser, {\it et al.}, Phys.Rev.Lett. {\bf 23},
880 (1969).

\bibitem{Walls} D.F. Walls and G.J. Milburn,
{\sl Quantum Optics} (Springer-Verlag, 1994).

\bibitem{jacobandmyung} H. Jeong, and M.S. Kim, Phys.Rev.A {\bf 65}, 042305
(2002).

\bibitem{deco} H. Jeong, J. Lee, and H. Nha,
J.Opt.Soc.Am.B, {\bf 25}, 1025 (2008).

\bibitem{referee} For ideal detectors and sufficiently large values of $d$ (for given values of $V$), 
our approach reproduces the results by Branciard {\it et al.} in \cite{Zeilinger07b} with a maximal 
violation of $0.125$ of Leggett's inequality at $\varphi\sim{0.25}$.

\bibitem{sqcat} A. Ourjoumtsev, {\it et al.}, Nature (London) {\bf 448}, 784 (2007).

\bibitem{strong} L. V. Hau, {\it et al.}, Nature (London) {\bf 397}, 594 (1999);
 P. Bermel, {\it et al.}, Phys.Rev.Lett. {\bf 99}, 053601 (2007).

\end{thebibliography}
\end{document}